\newif\ifarxiv
\arxivtrue

\ifarxiv
\documentclass[a4paper]{article}
\usepackage[top=1in, bottom=1.25in, left=1.25in, right=1.25in]{geometry}

\else
\documentclass[a4paper]{jpconf}

\fi

\usepackage{graphicx}
\usepackage{bm}        % for math
\usepackage{amssymb}   % for math
\usepackage{amsfonts}
\usepackage{amsmath}
\usepackage{epsfig}
\usepackage{units}
\usepackage{cite}
\usepackage[utf8]{inputenc}
\usepackage[T1]{fontenc}
\usepackage{color}
\usepackage{hyperref}
\newcommand{\pip}{$\pi^{+}$}

\newcommand{\kam}{K$^{-}$}

\newcommand{\Dzero}{\ensuremath{\mathrm{D^{0}}}}

\newcommand{\Dstar}{\ensuremath{\mathrm{D^{*\pm}}}}

\newcommand{\pp}{pp}

\newcommand{\GeVc}{GeV/$c$}
\newcommand{\GeVcsq}{GeV/$c^2$}

\newcommand{\s}{\ensuremath{\sqrt{s}}}

\newcommand{\pT}{\ensuremath{p_{\rm T}}}
\newcommand{\dedx}{d$E$/d$x$}

\newcommand{\ptchjet}{\ensuremath{p_{\mathrm{T,ch\, jet}}}}

\newcommand{\ptchjetgen}{\ensuremath{p_{\mathrm{T,ch\,jet}}^{\mathrm{part}}}}
\newcommand{\ptchjetdet}{\ensuremath{p_{\mathrm{T,ch\,jet}}^{\mathrm{det}}}}
\newcommand{\ptd}{\ensuremath{p_{\mathrm{T,D}}}}

\newcommand{\antikt}{anti-\ensuremath{k_{\mathrm{T}}}}

\ifarxiv

\title{Performance studies of D-meson tagged jets in pp collisions at $\s=7$~TeV with ALICE}
\date{}
\author{Salvatore Aiola, for the ALICE Collaboration \\
Physics Department, Yale University, New Haven, CT 06511\\
\href{mailto:salvatore.aiola@yale.edu}{salvatore.aiola@yale.edu}}
\begin{document}
\maketitle

\else

\begin{document}
\title{Performance studies of D-meson tagged jets in pp collisions at $\s=7$~TeV with ALICE}
\author{Salvatore Aiola, for the ALICE Collaboration}
\address{Physics Department, Yale University, New Haven, CT 06511}
\ead{\href{mailto:salvatore.aiola@yale.edu}{salvatore.aiola@yale.edu}}

\fi

\begin{abstract}
We present the current status of the measurement of jets that contain a D meson (D-tagged jets) with the \mbox{ALICE} detector.
\Dzero-meson candidates, identified via their hadronic decay into a K$\pi$ pair, were combined with the other charged tracks reconstructed with the central tracking system, 
using the anti-$k_{\rm T}$ jet-finding algorithm.
The yield of D-tagged jets was extracted through an invariant mass analysis of the D-meson candidates.
A Monte Carlo simulation was used to determine the detector performance and validate the signal extraction techniques.
\end{abstract}

\section{Introduction}
At hadron colliders, charm quarks are produced as a result of a hard scattering of partons. Like lighter quarks or gluons, charm quarks
fragment into collimated sprays of hadrons called \emph{jets}. The charm content of the jet is conserved throughout the fragmentation process,
which is dominated by Quantum Chromo-Dynamics (QCD).
In the final state, the charm content can be identified by looking for the presence of charmed hadrons among the jet constituents.

The measurement of the charm jet production cross section in \pp\ collisions is a sensitive test of perturbative QCD (pQCD) calculations~\cite{Cacciari:2012b}.
Heavy quarks are also an ideal probe of the Quark-Gluon Plasma (QGP)~\cite{STAR:2005a, PHENIX:2005a}
that is created in ultra-relativistic heavy-ion collisions. 
Hard scattered partons, including heavy quarks, interact with the QGP, which increases their virtuality and interferes with the
parton shower (\emph{jet quenching})~\cite{PHENIX:2008b, CMS:2012b, ALICE:2015a}.
At low parton energy, comparable to the charm mass, the charm quark is expected
to interact less strongly with the QGP~\cite{Dokshitzer:2001}, when compared to light quarks and gluons.

The production cross section of charmed hadrons has been measured
with good accuracy at the LHC~\cite{ALICE:2012d, LHCb:2013a, ATLAS:2016a, ALICE:2016b}.
Measuring the kinematic observables of jets with charm content implies integrating out some of the hadronization degrees of freedom. 
Since hadronization is a highly non-perturbative process, known only with large uncertainties~\cite{dEnterria:2014}, 
using observables less dependent on this process may improve comparisons with pQCD calculations.
Furthermore, the measurement of the fragmentation function (FF) of charmed hadrons 
can provide important insights into the charm production mechanism~\cite{CDF:1990, UA1:1990, STAR:2009a}.
ATLAS has measured the FF of \Dstar\ mesons, observing a large discrepancy with Monte Carlo
event generators~\cite{ATLAS:2012d}. ALICE has the potential to extend this measurement to low $z$ ($0.1<z<0.3$), 
where $z$ is the fraction of jet momentum carried by the D meson.

\section{The ALICE Experiment}
ALICE is the experiment dedicated to the study of heavy-ion collisions at the LHC.
The central barrel detectors ($\lvert \eta\rvert \lesssim 1$) are located inside a large solenoid magnet, providing a
field $B = 0.5$~T.
The \emph{Inner Tracking System} (ITS) is a six-layer silicon detector that allows a precise determination of the primary vertex 
and of displaced secondary vertices of weak decays.
The main tracking detector is the \emph{Time Projection Chamber} (TPC), which, combined with the ITS, allows reconstruction of tracks 
from low ($\pT\approx0.15$~\GeVc) to high transverse momentum
($\pT\approx100$~\GeVc) with good momentum resolution ($1-3\%$) and tracking efficiency ($60-80\%$).
Several detectors contribute to the Particle Identification (PID) capabilities of ALICE. 
In this analysis, we use the \dedx\ measured by the TPC and
the velocity $\beta$ measured by the \emph{Time Of Flight} detector,
consisting of an array of Multigap Resistive Plate Chambers.
A full description of the ALICE detector and of its performance during LHC Run-1 is available at Ref.~\cite{ALICE:2014b}.

\section{Analysis procedures}
The analysis relies on the well-established D-meson reconstruction techniques~\cite{ALICE:2012d, ALICE:2016a}, as well as
jet reconstruction methods~\cite{ALICE:2013c, ALICE:2015a, ALICE:2015e}, both developed by the ALICE Collaboration during Run-1.

For this study, only charged tracks were used to reconstructed the jets (\emph{charged jets}).
Track quality cuts were applied to ensure good momentum resolution.
The tightest cuts, which ensure the best possible momentum resolution at the expenses of tracking efficiency,
were applied to the tracks used to identify
the decay products of the D mesons.
In particular, at least one space point was required in one of the two layers of the ITS closest to the beam pipe.
For jet reconstruction, this last requirement was lifted to achieve a more uniform azimuthal efficiency.

\Dzero\ mesons ($m=1.865$~\GeVcsq, $c\tau=123\,\mu$m) and their charge conjugates were used to tag jets with charm content.
They were reconstructed via their hadronic decay: \Dzero $\rightarrow$ \pip \kam (BR = 3.88\%)~\cite{PDG:2016}. 
The topological cuts select unlike-sign (US) pairs that form a secondary vertex displaced from the reconstructed
primary vertex. PID on the \Dzero\ candidate daughters was used to reject pairs not compatible with the $\pi$K hypothesis.
The four-momenta of each \Dzero\ candidate and all reconstructed tracks
(excluding the two daughters of the \Dzero) were used as input to the \antikt\ jet-finding algorithm~\cite{Cacciari:2008c} with $R = 0.4$.
In view of replicating this analysis in background-rich environments, such as heavy-ion or high pile-up \pp\ collisions,
the \antikt\ algorithm was chosen for its good soft-resilient properties~\cite{Cacciari:2011a}.
Only jets containing the \Dzero\ candidates were retained.

In order to extract the signal out of the combinatorial background,
three different invariant mass analysis techniques were developed and their performance compared.
\begin{description}
\item[Invariant mass fit]
The invariant mass distribution of the \Dzero\ candidates was constructed for various intervals of their associated-jet transverse momentum (\ptchjet);
each candidate was given a weight corresponding to the inverse of its reconstruction efficiency $\epsilon(\ptd)$.
The invariant mass distribution was fit using the sum of an exponential function (background) and a Gaussian (signal). 
The yield $N^{\rm \Dzero\hbox{-}jet}(\ptchjet)$ was extracted from
the fit parameters.

\item[Side-band subtraction]
For each \ptd\ interval, the yield $N^{\rm \Dzero\hbox{-}jet}(\ptchjet,\ptd)$ was extracted by subtracting the
\ptchjet\ distribution in the side bands of the invariant mass distribution ($4\sigma_{\rm fit} < \lvert m - m_{\rm fit} \rvert < 8\sigma_{\rm fit}$) 
from the \ptchjet\ distribution in the peak area ($\lvert m - m_{\rm fit} \rvert < 2\sigma_{\rm fit}$); the peak position $m_{\rm fit}$, width $\sigma_{\rm fit}$ and side-band normalization factor were extracted 
by fitting the invariant mass distribution with an exponential + Gaussian function; the \ptd\ bins were weighted by the inverse of the \Dzero-meson reconstruction efficiency $\epsilon(\ptd)$ and summed over:
\begin{equation*}
N^{\rm \Dzero\hbox{-}jet}(\ptchjet)=\sum_{\ptd} \frac{1}{\epsilon(\ptd)} 
\left[N_{\rm peak}^{\rm \Dzero\hbox{-}jet}(\ptchjet,\ptd) - 
\frac{B_{\rm peak}^{\rm fit}}{B_{\rm SB}} 
N_{\rm SB}^{\rm \Dzero\hbox{-}jet}(\ptchjet,\ptd)\right],
\end{equation*}
where $B_{\rm peak}^{\rm fit}$ and $B_{\rm SB}$ are respectively the total background
in the peak area estimated by fitting the invariant mass distribution and the total
background in the side bands.

\item[Like-sign subtraction]
This method is analogous to the side-band method. In this case, the background \ptchjet\ distribution was provided by the like-sign (LS) $\pi$K pairs and
subtracted from the corresponding unlike-sign (US) pairs:
\begin{equation*}
N^{\rm \Dzero\hbox{-}jet}(\ptchjet)=\sum_{\ptd} \frac{1}{\epsilon(\ptd)} 
\left[N_{\rm US, peak}^{\rm \Dzero\hbox{-}jet}(\ptchjet,\ptd) - 
\frac{B_{\rm US, SB}}{B_{\rm LS, SB}} 
N_{\rm LS, peak}^{\rm \Dzero\hbox{-}jet}(\ptchjet,\ptd)\right],
\end{equation*}
where $B_{\rm US, SB}$ and $B_{\rm LS, SB}$ are respectively the integral of
the side bands in the US and in the LS invariant mass distributions.
\end{description}

Monte Carlo (MC) simulations were used to determine the detector performance and validate the analysis techniques.
PYTHIA6 (6.4.25)\cite{Sjostrand:2006} (tune Perugia-2011) was used 
to provide \pp\ collision events at $\s=7$~TeV.
Particles produced by PYTHIA were propagated through the detector using the GEANT3 transport code~\cite{GEANT3-url}.

\section{Results}
\label{sect:detperf}
\begin{figure}[tb]
\centering
\begin{minipage}{.48\textwidth}
\includegraphics[width=\textwidth]{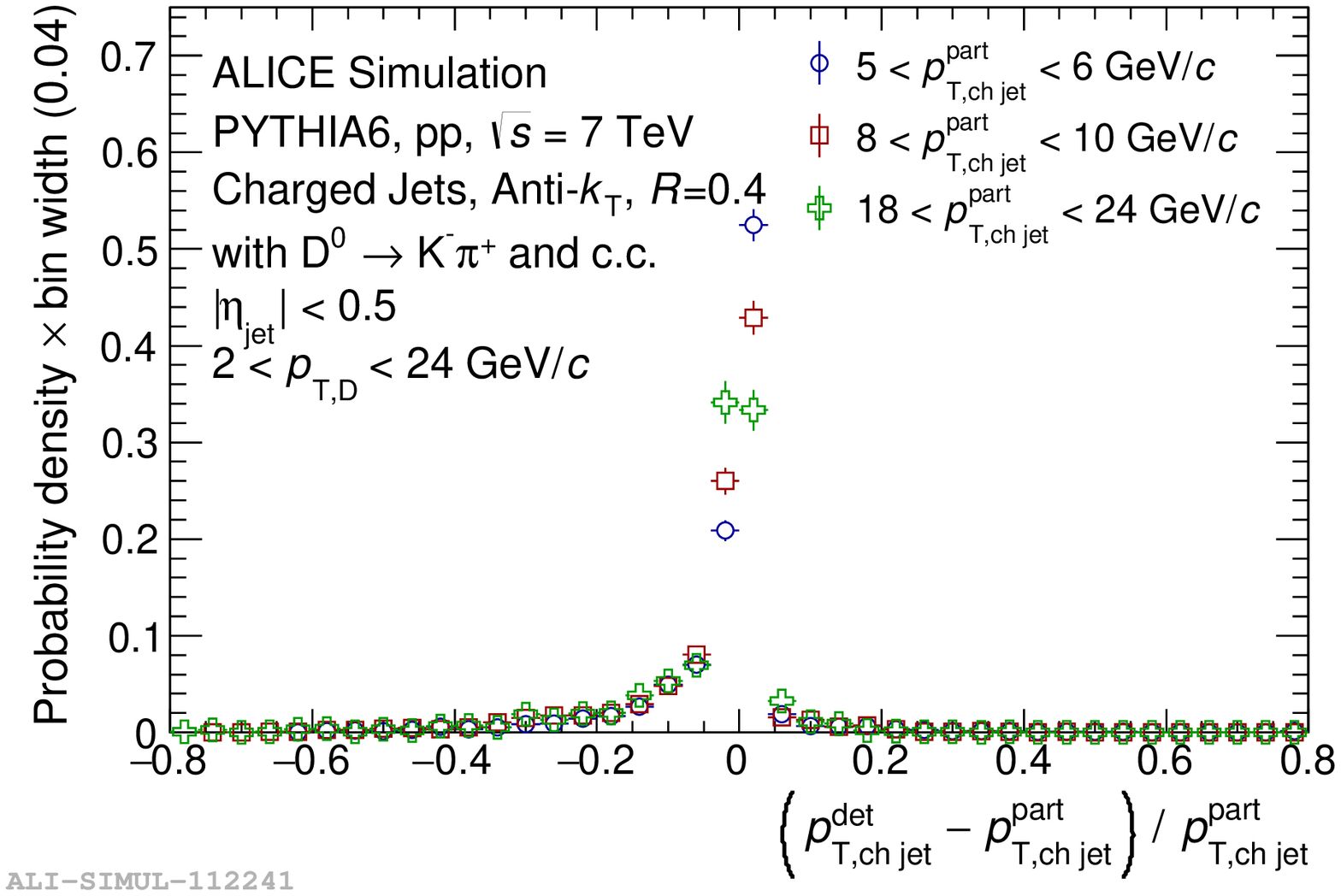}
\caption{\label{fig:HQ16_Simulation_DetectorResponse} Probability density distribution of the jet momentum shift in \ptchjet\ intervals.}
\end{minipage}\hspace{1pc}%
\begin{minipage}{.48\textwidth}
\includegraphics[width=\textwidth]{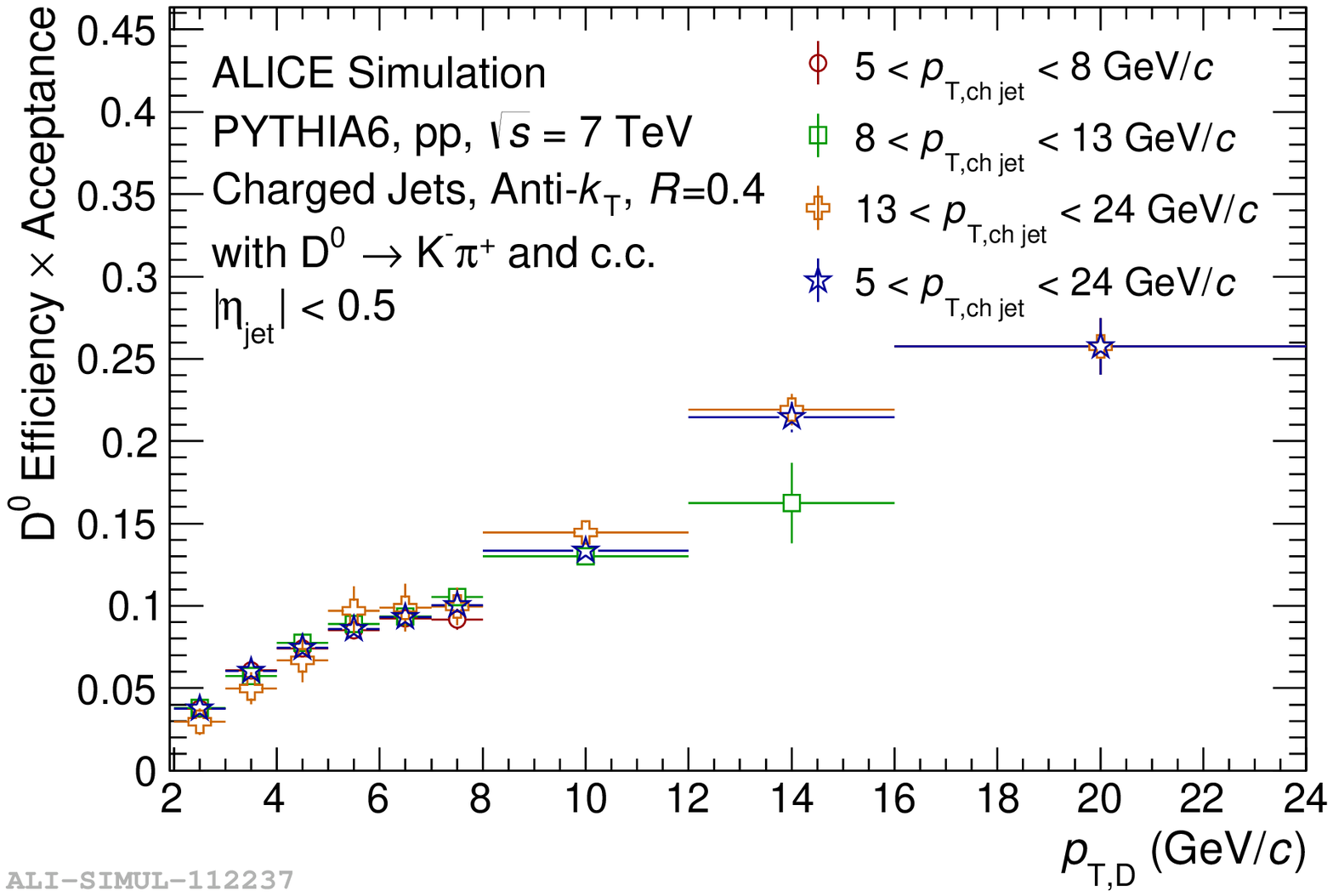}
\caption{\label{fig:HQ16_Simulation_EfficiencyVsDPt}Efficiency~$\times$~Acceptance of \Dzero\ mesons vs. \ptd\ in \ptchjet\ intervals.}
\end{minipage} 
\end{figure}
\ifarxiv
\begin{figure}[tb]
\centering
\includegraphics[width=.50\textwidth]{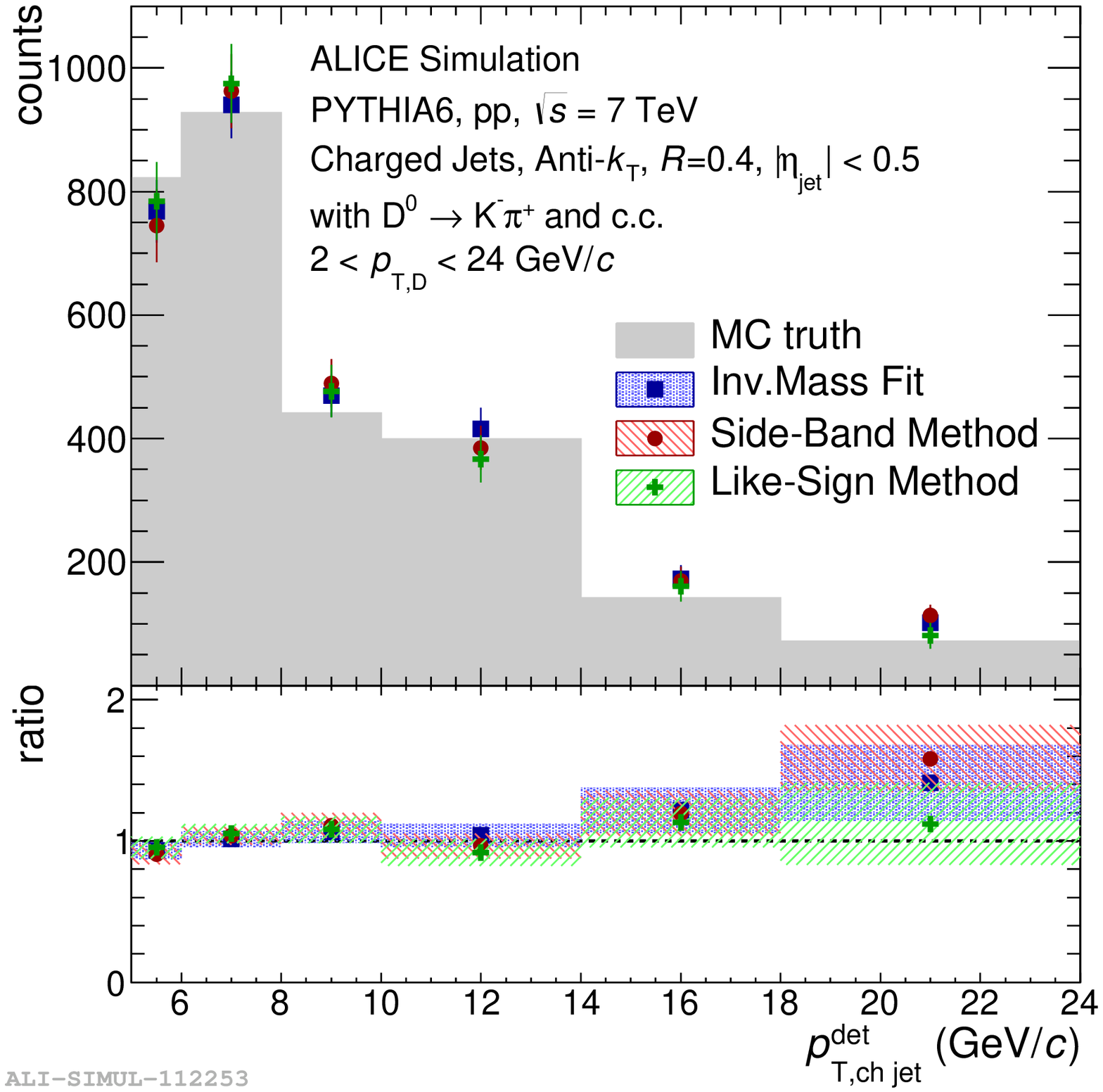}\hspace{1pc}%
\caption{\label{fig:HQ16_Simulation_MethodComparison}\Dzero-jet signal yield extracted using the invariant mass fit (blue squares), side-band method (red circles) and like-sign method (green triangles)
compared with the true D-tagged jet spectrum (gray filled area). The bottom panel shows the ratio to the MC truth. The yields are corrected for the reconstruction efficiency but not for distortions due to detector jet-momentum resolution (\mbox{$10-15\%$} on the yields).}
\end{figure}
\fi

In order to determine the detector performance, reconstructed detector-level D-tagged jets were matched with their corresponding counterparts at generator-level.
The variable
$\delta_{\ptchjet}=\left( \ptchjetdet - \ptchjetgen \right) / \ptchjetgen$,
was computed for each matched pair,
where \ptchjetdet\ and \ptchjetgen\ are the transverse momenta of the D-tagged jet at detector-level and at generator-level, respectively.
Its probability density distribution is shown in Fig.~\ref{fig:HQ16_Simulation_DetectorResponse} for three \ptchjetgen\ intervals. No significant dependence on \ptchjetgen\ was observed.
The shape of the distribution features a sharp peak at zero and is skewed towards negative values, due to tracking inefficiency (higher probability of
reconstructing smaller jet momenta than the generated ones). The jet momentum resolution (standard deviation of $\delta_{\ptchjet}$) for \Dzero-tagged jets is approximately \mbox{$11$\%}, 
slightly smaller with respect to its value for inclusive jets~\cite{ALICE:2015e}. The mean jet momentum shift is approximately \mbox{$-3$\%}.
The reconstruction efficiency is calculated as the ratio of the yield of reconstructed D-tagged jets over all generated D-tagged jets, as a function of generator-level observables.
It is shown in Fig.~\ref{fig:HQ16_Simulation_EfficiencyVsDPt} as a function of \ptd\ for different \ptchjet\ intervals; it shows a strong dependence on \ptd, mainly due to
varying topological cuts. No significant dependence on \ptchjet\ is observed in the interval $5<\ptchjet<24$~\GeVc.
Figure~\ref{fig:HQ16_Simulation_MethodComparison} shows the D-tagged jet yields as a function of \ptchjetdet\ obtained using the invariant mass fit, 
side-band and like-sign methods, compared to the true D-tagged jet spectrum. All signal extraction methods perform well and do not show significant biases, beyond the statistical uncertainties assigned to each.
\ifarxiv
\else
\begin{figure}[tb]
\includegraphics[width=.50\textwidth]{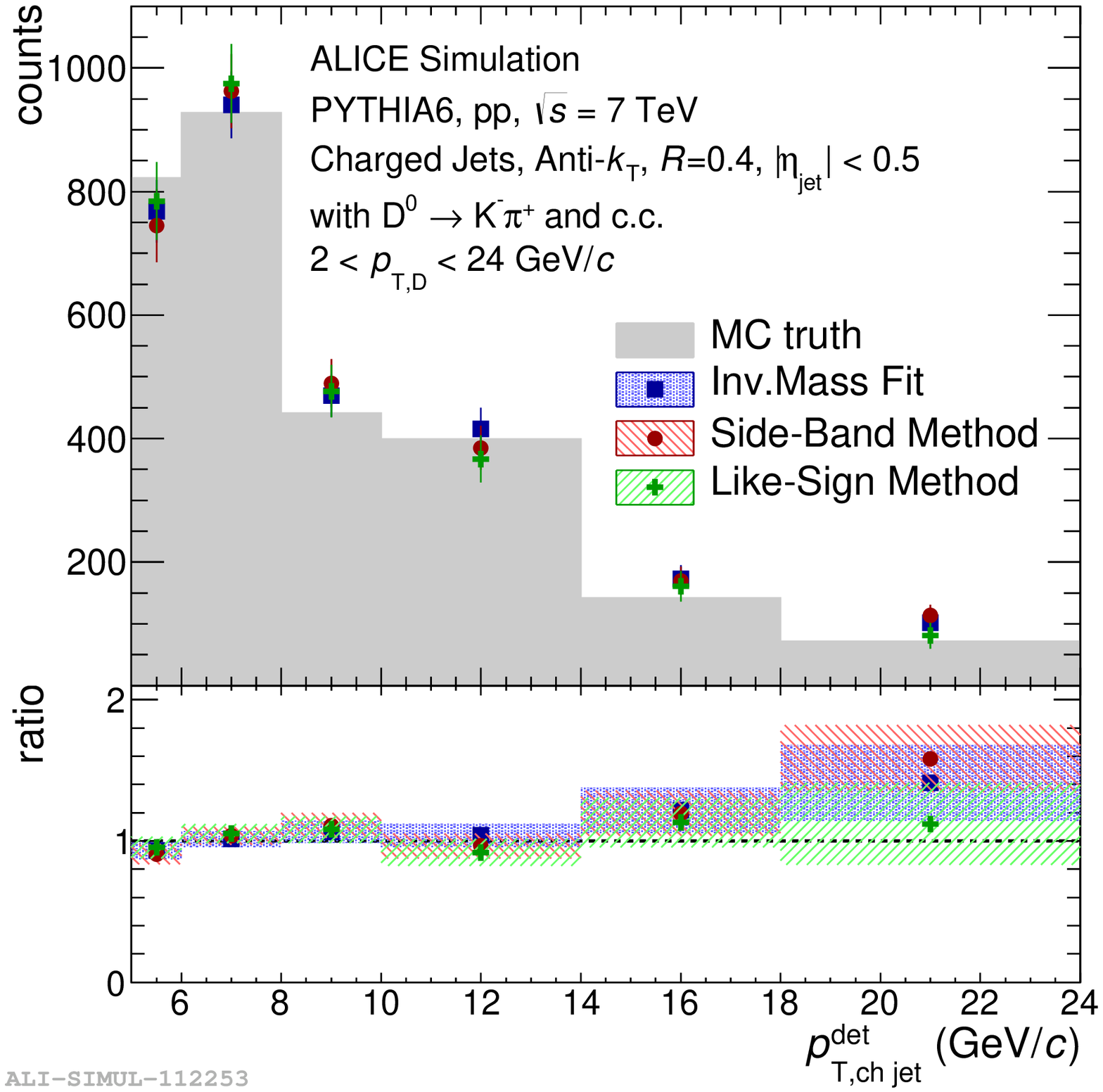}\hspace{1pc}%
\begin{minipage}[b]{.50\textwidth}\caption{\label{fig:HQ16_Simulation_MethodComparison}\Dzero-jet signal yield extracted using the invariant mass fit (blue squares), side-band method (red circles) and like-sign method (green triangles)
compared with the true D-tagged jet spectrum (gray filled area). The bottom panel shows the ratio to the MC truth. The yields are corrected for the reconstruction efficiency but not for distortions due to detector jet-momentum resolution (\mbox{$10-15\%$} on the yields).}
\end{minipage}
\end{figure}
\fi
\section{Conclusions}
ALICE has the potential for measuring jets with charm content tagged using reconstructed D mesons.
The ALICE detector performance for \Dzero-tagged jets was assessed using MC simulations
for pp collisions at $\s=7$~TeV. The \Dzero-jet momentum resolution was determined to be about 11\%; the reconstruction efficiency, independent of \ptchjet, ranges from 5\% to 25\% as a function of \ptd.
Three methods were implemented to extract the signal: invariant mass fit, side-band and like-sign subtraction.
The three methods were validated using a MC simulation and do not show biases larger than the statistical uncertainties.

\ifarxiv
\else
\section*{References}
\fi
\bibliography{biblio}{}
\bibliographystyle{iopart-num}

\end{document}